# Superior effect of edge relative to basal plane functionalization of graphene in enhancing polymer-graphene nanocomposite thermal conductivity – A combined molecular dynamics and Green's functions study

Rajmohan Muthaiah[a], Fatema Tarannum[a], Swapneel Danayat[a], Roshan Sameer Annam[a], Avinash Singh Nayal[a], N. Yedukondalu[b], Jivtesh Garg[a]

[a]School of Aerospace and Mechanical Engineering, University of Oklahoma, Norman,73019, USA

[b]Joint Photon Science Institute, Stony Brooks University, Stony Brook, New York 11794-2100, USA.

## ABSTRACT:

To achieve high thermal conductivity (*k*) of polymer graphene nanocomposites, it is critically important to achieve efficient thermal coupling between graphene and its surrounding polymers through effective functionalization schemes. In this work, we demonstrate that edge-functionalization of graphene nanoplatelets (GnPs) can enable a larger enhancement of effective thermal conductivity in polymer-graphene nanocomposites, relative to basal plane functionalization. Effective thermal conductivity for edge case is predicted, through molecular dynamics simulations, to be up to 48% higher relative to basal plane bonding for 35 wt.% graphene loading with 10 layers thick nanoplatelets. This unique result opens up promising new avenues for achieving high thermal-conductivity polymer materials, which is of key importance for a wide range of thermal management technologies. The anisotropy of thermal transport in single layer graphene leads to very high in-plane thermal conductivity (~2000 W/mK) compared to the low out-of-plane thermal conductivity (~10 W/mK). Likewise, in multilayer graphene nanoplatelet (GnP), the thermal conductivity across the layers is even lower due to the weak van der Waals bonding between each pair of layers. Edge functionalization couples the polymer chains to the high in-plane thermal conduction pathway of graphene, thus leading to high overall high composite thermal conductivity. Basal-plane functionalization, however, lowers the thermal resistance between the polymer and the surface graphene sheets of the nanoplatelet only, causing the heat conduction through inner layers to be less efficient, thus resulting in basal plane scheme to be outperformed by edge scheme. The present study fundamentally enables novel pathways for achieving high thermal-conductivity polymer composites.

**KEYWORDS:** graphene, functionalization, polymer nanocomposites, thermal conductivity, interface

## 1. INTRODUCTION

High thermal conductivity polymer materials can improve thermal management in a wide range of applications -such as water desalination[1], automotive control units[2], batteries[3], solar panels[4], solar water heating[5], electronic packaging[6], and electronic cooling[7]. A key approach to enhance thermal conductivity of polymers is addition of high thermal conductivity fillers such as graphene ($k$ ~ 2000 W/mK)[8-10]. The success of this approach is, however, limited by the large interface thermal resistance between graphene and polymer in the range of $10^{-8}$ to $10^{-7}$ $m^2$ $KW^{-1}$ [11,12] due to mismatch of phonons (lattice vibrations) between these two. To decrease thermal interface resistance, graphene is chemically functionalized by groups that are compatible with the surrounding polymer[21]. Recent work demonstrated that multilayer graphene is more efficient at enhancing thermal conductivity than single layer graphene[13]. For such multilayer graphene, it is critically important to understand the optimal location of functional groups (such as edge or basal plane of graphene) which can lead to lowest interface thermal resistance. Different location of functional groups can generate vastly different enhancements in overall polymer thermal conductivity because of the differences in resulting thermal coupling of the nanoplatelet with the polymer. In this work we demonstrate for the first time, that functionalization on the edges can lead to significantly higher effective polymer nanocomposite thermal conductivity compared to functionalization on the basal plane. Detailed understanding of the effect is achieved through molecular dynamics as well as first-principles simulations based on density-functional theory. Presented results unravel promising new avenues for achieving high thermal conductivity polymer-graphene nanocomposites.

This benefit of edge-bonding is found to be related to enabling all graphene sheets to be covalently bonded to surrounding polymer chains, as the edges of all sheets are exposed to surrounding polymer (Fig. 1a) High in-plane thermal conductivity of individual sheets (~2000 W/mK)[8,9] combined with their efficient coupling with the surrounding polymer through edge functionalization leads to efficient thermal conduction pathway across the GnPs (Fig. 1a). Since all sheets are efficiently coupled with polymer through edge bonding, the entire nanoplatelet is an efficient in conducting heat through edge functionalization (Fig. 1a). Basal plane bonding, on the other hand, primarily couples only the outermost surface layers of the nanoplatelet (Fig. 1b). The weak van der Waals coupling of outer layers with inner layers renders the inner layers to be less efficient in heat transfer due to poor through-thickness thermal conductivity of graphene (~10 W/mK)[12,14] (Fig. 1b). The lower heat transfer capability of inner layers for basal plane bonding, causes the overall nanoplatelet heat conduction to be lower for basal plane bonding.

Several studies have been reported enhancement of thermal conductivity through functionalization. Two orders of magnitude increase in interface thermal conductance[22] and 156% enhancement[23] in composite thermal conductivity were achieved through grafting of polymer chains on to graphene. Theoretical studies on polymer grafted graphene (performed by Wang et al.) showed two-fold lower interfacial thermal resistance through attachment of functional groups[15]. Also, pyrene-end poly(glycidyl methacrylate) functionalized graphene/epoxy composite achieved ~184% enhancement due to noncovalent functionalization[16]. Furthermore, Konatham et al.[17] demonstrated almost 50% reduction in the interfacial thermal resistance between functionalized graphene and octane using molecular dynamics (MD) simulations study. Lin et al.[18] experimentally explored the enhancement of out-of plane thermal conductance of graphene facilitated by functional group of alkyl-pyrene on graphene via MD simulations. According to Ganguli et al.[19] silane-functionalized graphene improved the thermal conductivity by 50% compared to pristine graphene composite for 8% filler content. Yang et al.[20] described the comparison of the interfacial tension behavior of edge and basal plane functionalized graphene at both liquid−liquid and liquid−solid interfaces. Mungse et al.[21] studied the lubrication potential of basal plane functionalized alkylated graphene nanosheets. In another study, the effect of interconnected 3D network structure of edge or basal plane modified graphene oxide on electrical conductivity[22] was examined. Xiang et al.[23] also compared edge versus basal plane functionalization of graphene for energy conversion and energy storage applications. However, there is a lack of detailed understanding of the relative effectiveness of edge versus basal plane functionalization in enhancing thermal conductivity of polymer-graphene nanocomposites. This study addresses the issue with combined molecular dynamics simulation and first principles driven atomistic Green's function methods.

In this work, detailed molecular dynamics simulations are performed to predict and understand the effects of edge and basal plane functionalized graphene on thermal conductivity enhancement of the polymer nanocomposites. Computational studies are also performed to understand the beneficial effects of edge bonding on the: (a) thermal conductivity of individual graphene sheets (using MD simulations); (b) interface thermal conductance at individual junctions between graphene and polymer (performed using first principles driven atomistic Green's function methods); and (c) damping of vibrations in the outer layers of a graphene nanoplatelet.

## 2. METHODS

### 2.1 Molecular Dynamics Simulations

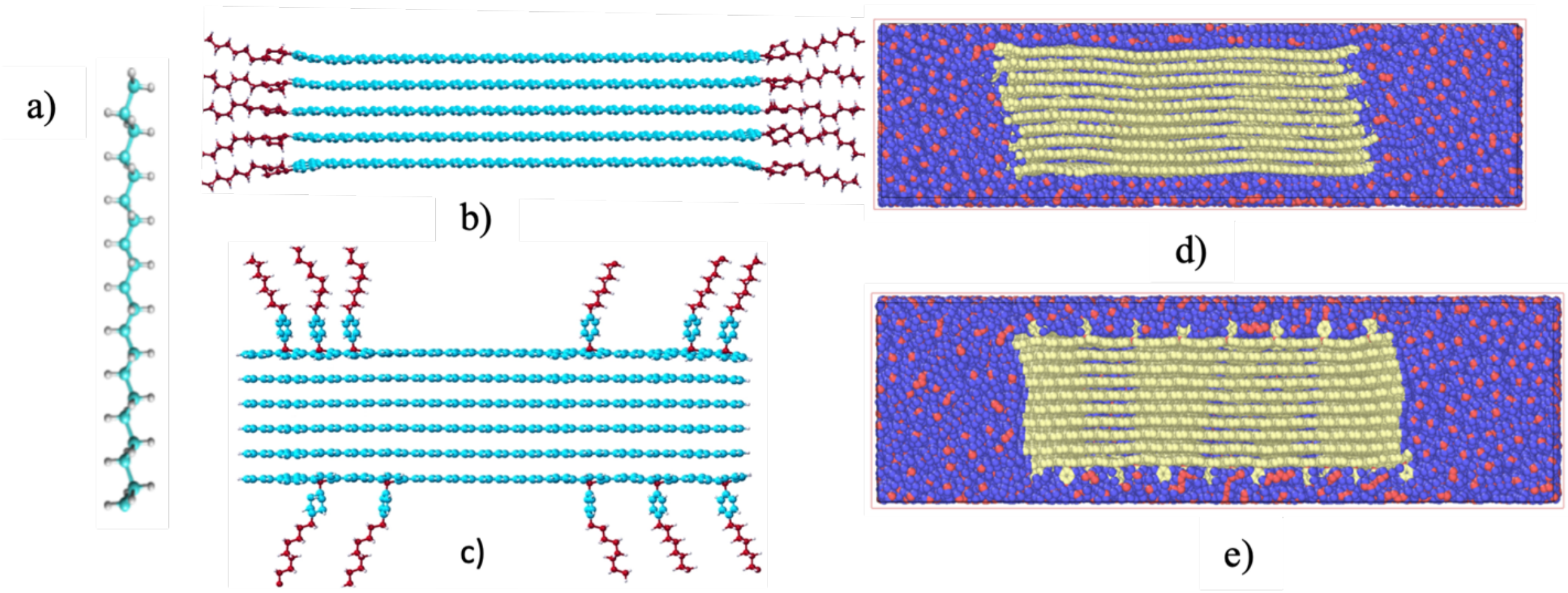

**Fig. 1a) Polyethylene chain with 120 atoms b and c) Edge and basal-plane functionalized graphene nanoplatelet d and e) Edge and basal plane functionalized graphene nanoplate embedded in polyethylene**

Molecular dynamics simulations were performed using LAMMPS package[24]. Interatomic force interactions between various atomic species were modeled using COMPASS force-field[25]. This force field has been widely used in the past to simulate polymer systems[25]. Polymer used for computational studies is polyethylene (PE). To compare edge and basal plane case computationally, we first functionalized the graphene nanoplatelet separately on its edges and basal plane to achieve edge (Fig. 1b) and basal plane functionalized (Fig. 1c) nanoplatelet. These functionalized nanoplatelets were then embedded into the polyethylene matrix (using PACKMOL package[26]) to yield edge functionalized (Fig. 1d) and basal plane functionalized (Fig. 1e) polyethylene-graphene nanocomposites. The lateral dimension of the GnPs used for simulations was 10 nm x 10 nm while its thickness was varied from 4 to 10 layers. The polymer-graphene composite was prepared with a graphene nanoplatelet concentration of 35 weight%. The chain length of polyethylene used was 120 Carbon atoms. The system was relaxed using NPT ensemble (constant number of particles, pressure and temperature) to achieve an equilibrium configuration. Simulations were performed at 300 K. A temperature difference of 50 K was imposed across the composite (using Langevin thermostats[27]) and simulations were performed until steady state was reached (typically after 5 ns). We studied a Langevin thermostat between 0.4 nm and 1.2 nm and observed a convergence in heat flux after 0.6 nm. Hence, we maintained 0.6 nm Langevin thermostat for all the calculations. Upon reaching steady state the resulting heat flux was computed and compared for edge and basal plane functionalization cases. Energy exchange between the thermostated regions was plotted with time as shown in Fig 1. The heat flux was calculated from the slope of the "energy exchange versus time" graph using the equation, $J=\Delta E/(A.\Delta t)$, where $\Delta E$ is the change in energy, A is the cross-sectional area and $\Delta t$ is the time over which $\Delta E$ is computed. These calculations were performed using the linear portion of the graph (which corresponds to steady state). Energy exchange ($\Delta E$) with time is shown in Fig 3e for composites prepared with edge functionalized (EFGNP) and basal plane functionalized (BFGNP) graphene nanoplatelet.

To validate our models, we simulated the thermal conductivity of pure polyethylene with a density of 0.898 $gcm^{-3}$ and obtained a value of 0.3 W/mK at 300 K which is in good agreement with simulations[28-31] and experimental results[32,33].

## 2.2 Atomistic Green's Function (AGF) Method

A typical AGF simulation system for phonon transmission consists of three regions, the interfacial region and two semi-infinite reservoirs (left and right) made up of bulk crystal 1 and 2 (Fig. 2). Phonon transmission across the interfacial region is calculated using Green's functions, which provide response of the system under a small perturbation. Under harmonic approximation, the Green's function G—corresponding to the interfacial region—can be calculated as[34], $G_{d,d} = [\omega^2 I - H_{d,d} - \Sigma_R - \Sigma_L]^{-1}$, where $\omega$ is the phonon frequency, $H_{d,d}$ represents the dynamical matrix of the interfacial region, and $\Sigma_L$ and $\Sigma_R$ are the self-energies of the left and right reservoirs. Self-energies represent the effect of contact reservoirs on the interfacial region. With the knowledge of Green's function, the total phonon transmission across the interfacial region is calculated as $\Xi(\omega) = Trace[\Gamma_L G_{d,d} \Gamma_R G_{d,d}^{+}]$ where $\Gamma_R$ and $\Gamma_L$ describe the rate at which phonons enter and exit the right and left contact leads respectively, $\Gamma_L = i[\Sigma_L - \Sigma_L^+]$, $\Gamma_R = i[\Sigma_R - \Sigma_R^+]$, and '+' denotes the conjugate transpose of the matrix. The interface thermal conductance can then be calculated with Landauer formalism using the total phonon transmission,

$$J = \int \left(\frac{\hbar\omega}{2\pi}\right) \Xi(\omega) \frac{dN(\omega)}{dT} \, d\omega$$

where $N(\omega)$ and $T$ are the Bose-Einstein population and temperature respectively.

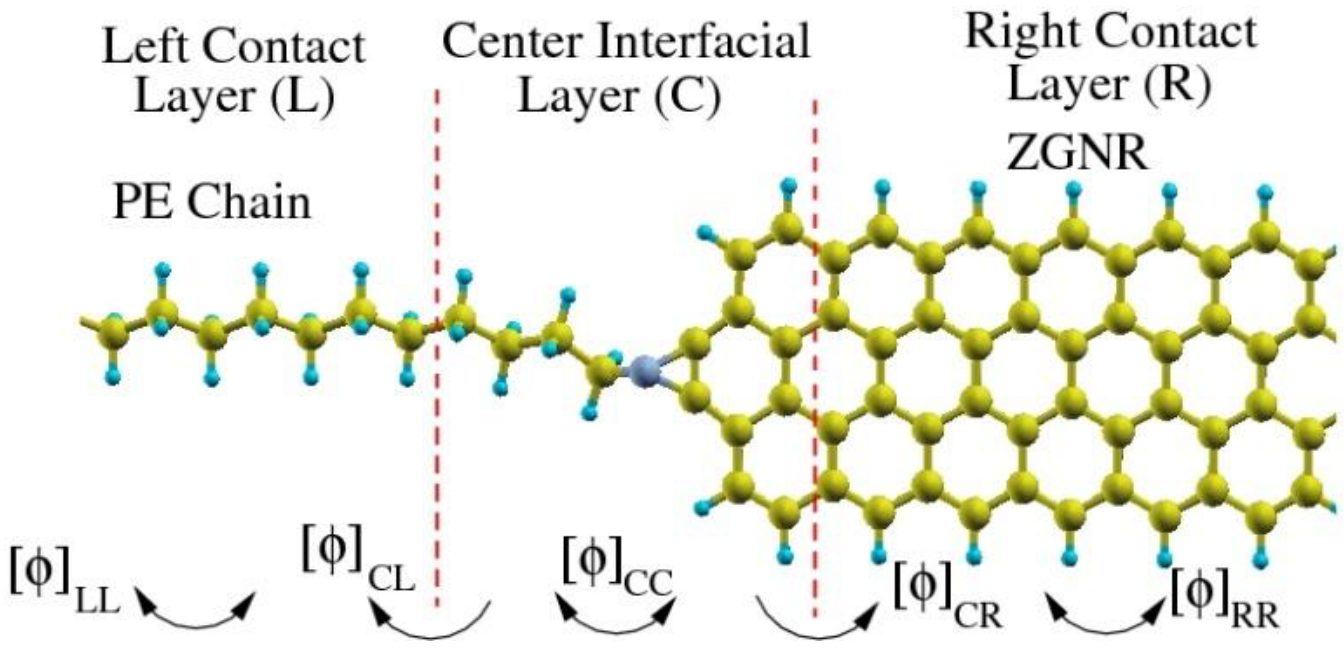

**Fig. 2:** Atomic configuration for atomistic Green's function calculation of interfacial thermal conductance.

The only inputs needed to compute the various quantities involved in calculation of thermal conductance are the harmonic interatomic force constant (IFC) matrices $[\phi]$ (as shown in Fig. 2) and the masses of atoms in the various layers. The harmonic force constant matrix $[\phi_{ij}]_{\alpha\beta}$, between two regions $\alpha$ and $\beta$ ($\alpha$, $\beta$= L, R or C) is comprised of the interatomic force interaction $\phi_{ij}$ between any atom $i$ in region $\alpha$ and another atom $j$ in region $\beta$. These force constants $\phi_{ij}$ are the second derivatives of energy with respect to the displacements of atoms $i$ and $j$. In this work these IFCs are derived accurately from density-functional theory (DFT) using open source package Quantum Espresso[35]. Use of DFT has been shown to yield very accurate IFCs in prior efforts[36]. In our calculations, the right contact layer is the graphene nanoribbon and the left contact layer is a single chain of polyethylene. The center layer consists of the interfacial junction. Ultrasoft pseudopotentials were used for performing electronic calculations. An energy cutoff of 60 Ry was used for wavefunctions, and 480 Ry was used for charge density.

To understand the higher phonon transmission for edge case, we split the total transmission in terms of transmission from polymer to out-of-plane and in-plane vibration modes in graphene. This is accomplished by an extension of AGF method to compute transmissions for individual phonon branches[37]. To achieve such decomposition, it should be noted that matrices $\Gamma_L$ and $\Gamma_R$ can be written as $\Gamma_L = \phi_{LC} A_L \phi_{CL}$ and $\Gamma_R = \phi_{RC} A_R \phi_{CR}$, where $\phi$ are the force interaction matrices as discussed earlier. In these expressions matrix $A$ is proportional to phonon density of states and can be written in terms of its eigenvectors and eigenvalues as[37] $A = \sum_i \lambda_i e_i \, e_i^+$ , where $\lambda_i$ and $e_i$ are the eignevalues and eignevectors of matrix $A$ respectively[37]. Above decomposition of matrix A allows $\Gamma_L$ and $\Gamma_R$ to be replaced with polarization dependent phonon escape rates defined as $\gamma_L = \phi_{LC}\lambda_{L,i}e_{L,i}e_{L,i}^{+}\phi_{CL}$ , $\gamma_R = \phi_{RC}\lambda_{R,i}e_{R,i}e_{R,i}^{+}\phi_{CR}$. This allows phonon transmission from left into specific phonon polarizations on the right to be written as $\xi(\omega) = Trace[\Gamma_L G_{d,d}\gamma_R G_{d,d}^{+}]$. In this work, we take the phonon escape rates for out-of-plane and in-plane vibration modes in the right contact to be

$$\gamma_R^{out} = \phi_{RC}\left[\sum_{i,i\in out}\lambda_{R,i}e_{R,i}e_{R,i}^{+}\right]\phi_{CR}$$

$$\gamma_R^{in} = \phi_{RC}\left[\sum_{i,i\in in}\lambda_{R,i}e_{R,i}e_{R,i}^{+}\right]\phi_{CR}$$

The transmission from polymer in to the out-of-plane and in-plane vibration modes in graphene can then be computed as $\xi^{out}(\omega) = Trace[\Gamma_L G_{d,d}\gamma_R^{out} G_{d,d}^{+}]$ and $\xi^{in}(\omega) = Trace[\Gamma_L G_{d,d}\gamma_R^{in} G_{d,d}^{+}]$.

## 2. RESULTS AND DISCUSSION

### 2.1 Molecular Dynamics Simulations

The heat flux was calculated from the slope of the "energy exchange versus time" graph using the equation, J=ΔE/(A.Δt), where ΔE is the change in energy, A is the cross-sectional area and Δt is the time over which ΔE is computed. These calculations were performed using the linear portion of the graph (which corresponds to steady state). Energy exchange (ΔE) with time is shown in Fig 3e for composites prepared with edge functionalized (EFGNP) and basal-plane functionalized (BFGNP) graphene nanoplatelet. Results of heat flux computation show a significant increase of nearly 48% for heat flux for the EFGNP/PE nanocomposite case over the BFGNP/PE nanocomposite case for the 10 layers thick nanoplatelet (Fig. 3c). The enhancement in heat flux is found to be thickness dependent (Fig. 3d), with thicker nanoplatelets demonstrating larger advantage of edge relative to basal plane functionalized cases. This is seen in Fig. 3d, where it is visible that while for 4 sheets thick nanoplatelet, EFGNP/PE composite has 25% higher heat flux compared to BFGNP/PE composite, increasing the sheet thickness to 10 layers, increases the difference in heat flux to 48%. These data provide first computational understanding of the superior effect of edge-bonding in enhancing polymer-graphene nanocomposite thermal conductivity.

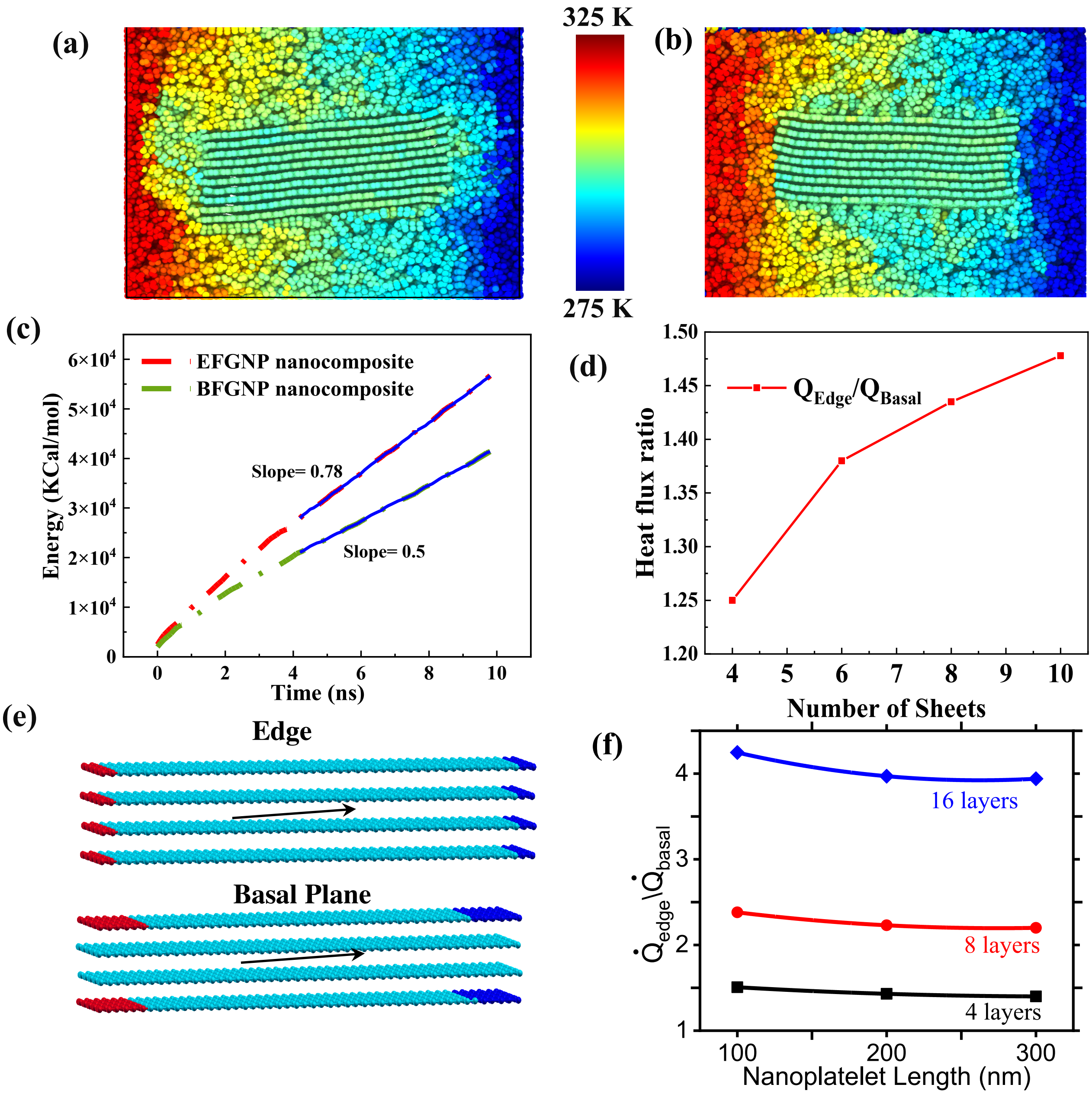

**Figure 3.** (a) EFGNP/PE and (b) BFGNP/PE nanocomposites. (c) Energy transfer vs time for edge and basal plane functionalized nanocomposites and (d) heat flux ratio for edge vs basal functionalized nanocomposite as function of nanoplatelet thickness. (e) Temperature boundary conditions to simulate heat flux through single nanoplatelet for edge and basal plane functionalization (arrows show direction of heat flow) (f) Heat flux ratio of different length of nanoplatelet shows higher heat flux for edge bonding.

This higher heat flux is mainly due to all layers of graphene nanoplatelet being efficiently coupled to polymer matrix leading to lower interfacial thermal resistance for edge bonding case (Fig. 1b). For basal plane case, however, only the outermost layers are the dominant heat conductors, as they are covalently bonded with the surrounding polymer (Fig. 1c). For this case, inner layers interact with the outer layers through weak van der Waals forces, diminishing heat transfer from outer to inner layers, thus diminishing the contribution of inner layers to overall heat transfer for basal plane case. This degrades the heat transfer performance for basal plane functionalized graphene nanoplatelet. These differences in heat transfer reflect in the temperature profiles across the nanocomposite for the two cases; for the edge case, temperature profile is significantly smoother at the interface (at the edge) between graphene and polymer (Fig. 3a), relative to basal plane case (Fig. 3b), where a sudden temperature jump is seen across the nanoplatelet edge. This indicates a smaller interface thermal resistance for edge functionalization case, resulting in higher effective thermal conductivity.

While above simulations study the entire nanocomposite, we also performed simulations for single nanoplatelet (Figs. 3e and 3f) to demonstrate the effect more clearly. Fig. 3e shows single graphene nanoplatelet with the thermostats applied across either the edges of the entire nanoplatelet (to simulate edge bonding) or across only the outermost layers (to simulate basal plane bonding). These thermostats establish a temperature gradient of 50 K. Fig. 3f shows comparison of heat transfer rate between edge and basal plane case. Edge case is seen to outperform basal plane case as seen by the ratio ($\dot{Q}_{edge}/\dot{Q}_{basal}$) being greater than 1.

It is further seen that the advantage of edge bonding (indicated by the heat transfer rate ratio) increases with increasing number of layers (*n*) within the nanoplatelet. While for a 4-layer thick nanoplatelet, the heat transfer rate ratio is ~1.5, this ratio increases sharply to more than 4 for a 16-layer thick nanoplatelet. The increase in difference between edge and basal plane functionalized case with increase in $n$ can be understood by realizing that for the basal plane functionalized nanoplatelet, increasing the number of inner layers does not result in significant increase in overall heat transfer rate, since inner layers do not conduct heat efficiently due to being poorly coupled with outermost layers (which receive heat through functionalization). For edge case, however, since all layers conduct heat efficiently, increase in $n$ leads to a proportional increase in overall heat transfer. This causes the ratio of heat transfer for edge and basal plane functionalization to increase with number of graphene sheets. This also explains the increase in heat flux ratio for edge and basal plane cases with increase in thickness for the nanocomposite (Fig. 3d). Fig. 4 shows the effect of grafting density (0.25% - 1%) and number of backbone carbon atoms in the functionalized chain. It is observed that, the difference in heat flux increases with increasing in both grafting density and number of carbon atoms and saturates at 1% which is in good agreement with the previous work[38,39]. At high higher number of backbone carbon atoms, a functionalized molecules bend, fold and twine itself and increases the interface thermal resistance[39]. Hence, we observed a maximum difference in heat flux of 48% is observed at 1% grafting density and 12 backbone carbon atoms.

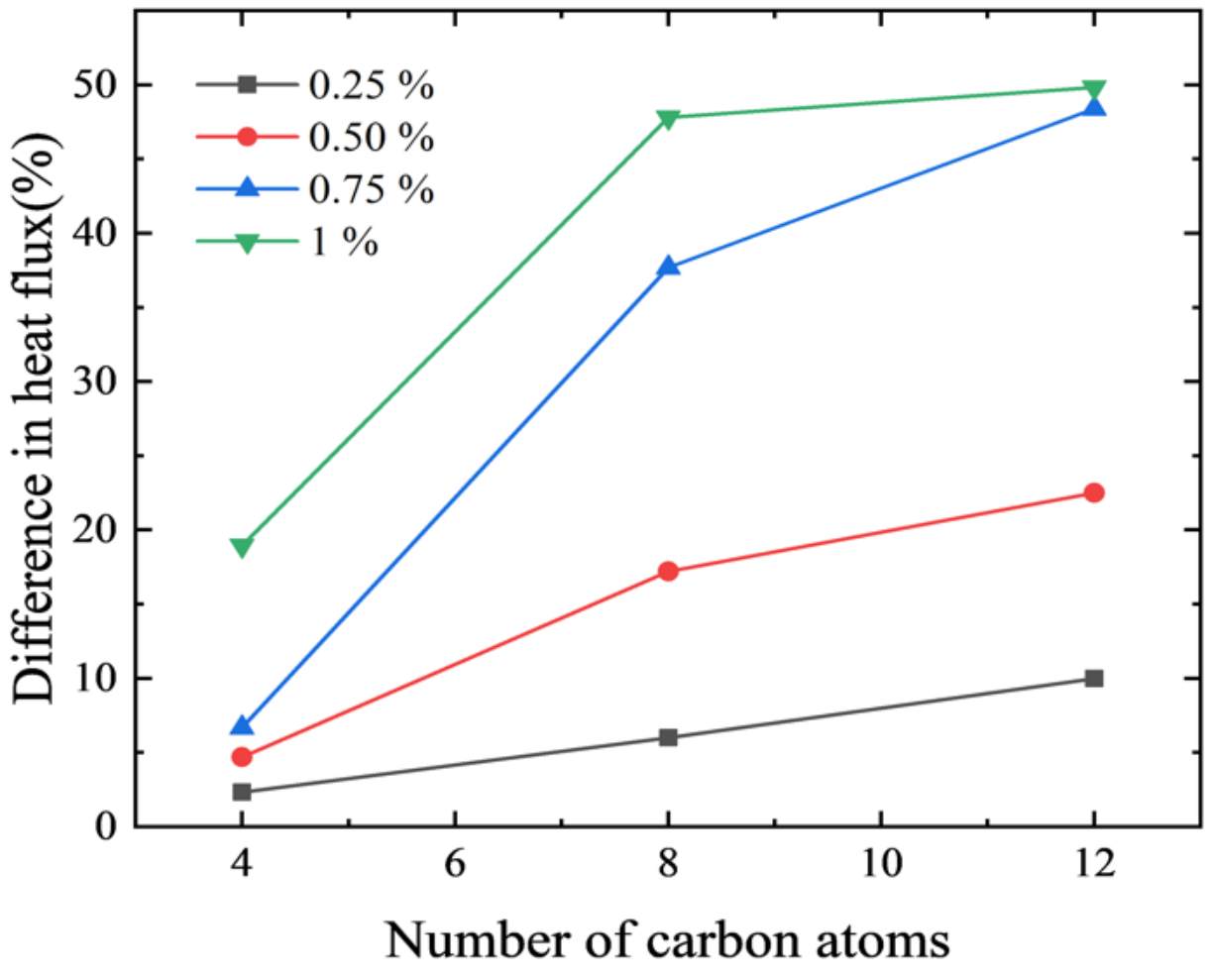

Figure 4: Difference in heat flux with grafting density and number of backbone carbon atoms in the functionalized chain.

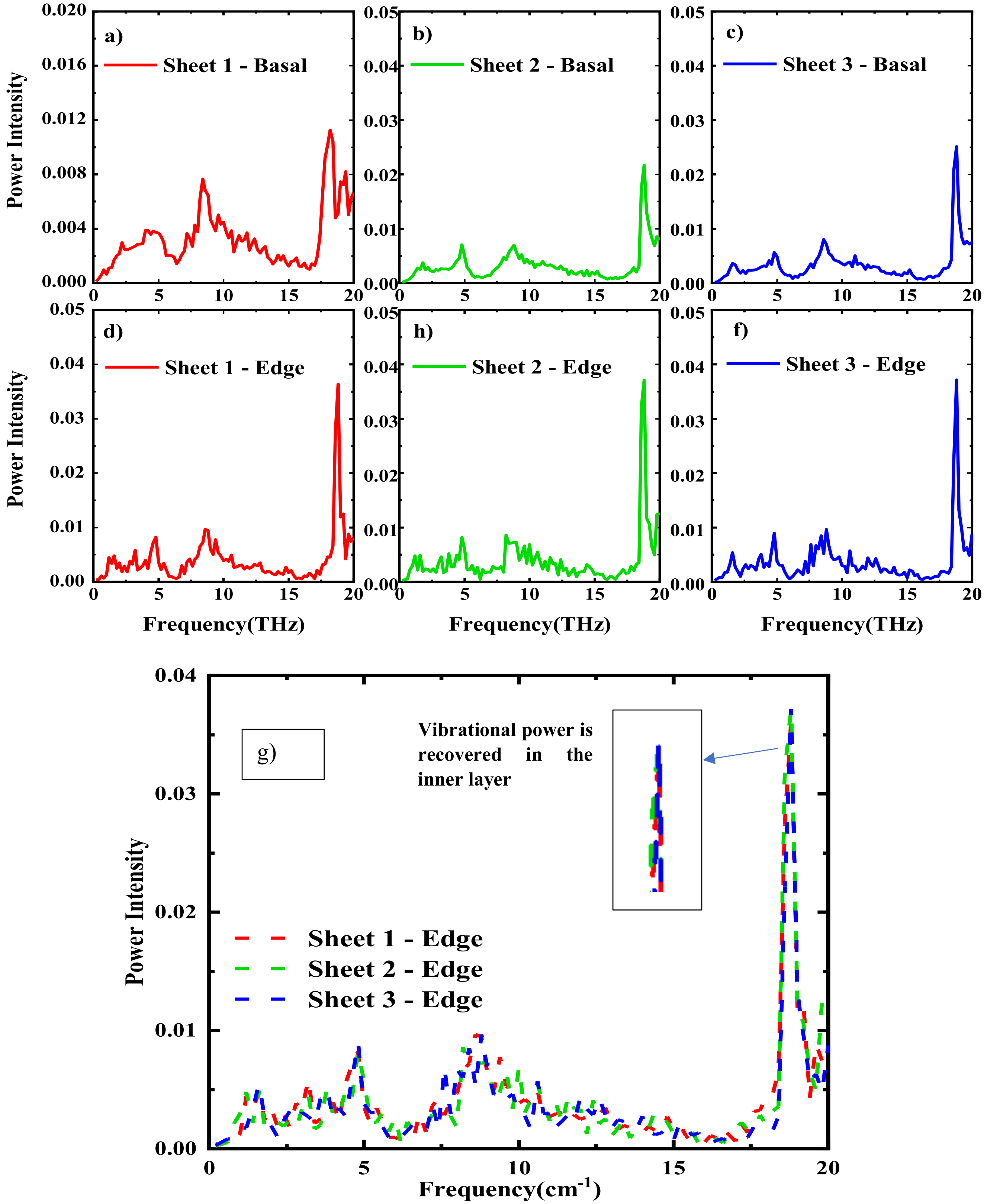

Figure 5. (a-g): Vibrational power spectra across different layers in EFGNP and BFGNP. Vibrational power in EFGNP inner layer is higher than BFGNP due to strong coupling of inner layers with the polymer for EFGNP.

In this work, edge bonding not only enables the coupling effect of all graphene layers to polymer, but also brings out other advantages, namely, a) lower damping of vibrations in graphene layers by surrounding polymer for edge bonding case, b) higher thermal conductivity of individual edge-functionalized graphene sheets (studied using MD), c) higher interface thermal conductance at an individual junction between polymer and graphene on the edge compared to basal plane (studied using first-principles atomistic Green's function analysis).

We first discuss effect of edge bonding on damping of vibrations within graphene sheets. Vibrational power density spectrum[11] is a powerful method to study damping of vibrations in graphene sheets and is computed from the discrete Fourier transform of the velocity autocorrelation function as shown below,

$$D(\omega) = \int_0^{\tau} < v(0).v(t) > \exp(-i\omega t)dt \qquad (1)$$

where $< v(0).v(t) >$ is the velocity autocorrelation obtained by correlating the velocity at every 2 fs time interval, $\tau$ is the total correlation time = 5 ps and $D(\omega)$ is the phonon vibrational power spectra at frequency$(\omega)$.

Figs 5a-h shows the difference in vibrational power spectra for edge and basal plane functionalization cases for a 6 sheet nanoplatelet (embedded in the polymer matrix). The power spectra are presented for individual sheets within the nanoplatelet. Sheet 1 denotes the outermost layer from bottom. We have shown the vibrational power spectra of sheets 1-3. We focus on vibrational power density at a frequency of around 18 THz[40]. Comparing Fig. 5 a and d, it is seen that vibrational power in Sheet 1 (outermost sheet) is lower in BFGNP than EFGNP by almost a factor of 3.5. This stronger damping of vibrations in outermost sheet for BFGNP case is caused by strong covalent-bond mediated interaction between outer sheet and polymer for BFGNP case, in contrast to the case of edge bonding where the interactions are much weaker caused by van der Walls forces.

Vibrational power for inner sheets is also higher for edge case compared to basal plane functionalization. The inner layers for edge case do not have a large contact area with the surrounding polymers (only interact through edges) and therefore are less damped compared to outer sheet, causing a small increase in their power density (see inset in Fig. 5g) relative to outer sheet. The inner sheets for edge case at the same time receive heat from edges through strong covalent bonds. For basal plane case, however, vibrational power for inner sheets stays lower relative to edge case. This is simply due to the weak van der Waals interaction of inner layer with the outer layer for basal plane case which causes poor heat transfer to the inner layers for basal plane case. Above comparison of vibrational power density spectrum highlights the significant advantages of edge bonding for heat conduction.

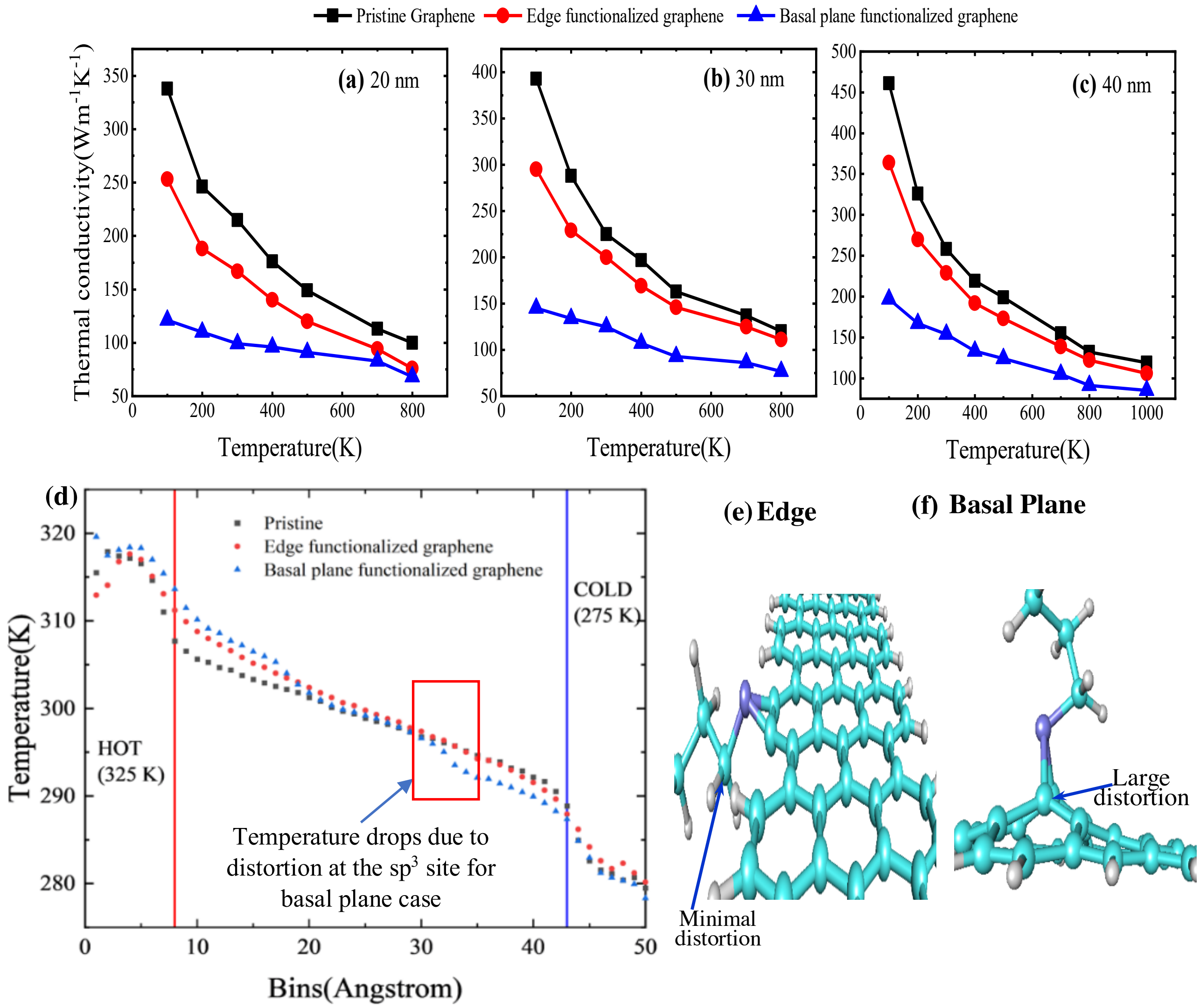

**Figure 6.** Temperature dependence of thermal conductivity for pristine, edge functionalized and basal plane functionalized graphene at (a) 20 nm, (b) 30 nm, and (c) 40 nm lengths (d) Temperature distribution along the simulation box in pristine, edge, and basal plane functionalized graphene. Relaxed atomic configurations for (e) edge and (f) basal plane functionalization demonstrates larger distortion of graphene for basal plane functionalization.

We next compare the effect of edge and basal plane bonding on thermal conductivity of individual graphene sheets. Edge-bonding is found to enable remarkably superior thermal conductivity of individual graphene sheets relative to basal plane functionalization (Fig. 6). For pristine graphene case, *k* at nanometer region is found in good agreement with the previous work[41]. At length scale of 20 nm and at room temperature (300 K), the thermal conductivity (*k*) of pristine, edge functionalized, and basal plane functionalized graphene are computed to be 215 W/mK, 167 W/mK, and 99 W/mK, respectively. The thermal conductivity predicted for edge-functionalized case is 68.7% higher than the basal plane case and 28.7% lower than pristine graphene. This reduction in thermal conductivity upon functionalization (relative to pristine graphene) is due to the distortion of the graphene structure. Large reduction in thermal conductivity in basal plane functionalization is attributed to much larger distortion of graphene in its basal plane as shown in Fig 6f. Edge bonding distorts graphene to a much smaller degree compared to basal plane case, as seen in the relaxed DFT (density-functional theory) structures in Fig. 6e and f. Carbon atoms on the basal plane of graphene are $sp^2$ hybridized; in forming an extra bond to functionalize, they transform to $sp^3$ state, protruding outwards and distorting graphene in the process. Unlike inner carbon atoms, edge atoms can adopt tetrahedral geometries more freely without causing extra strain. Lower distortion through edge bonding results in significantly higher $k_{graphene}$ relative to the basal plane case. To demonstrate this effect clearly, we present the temperature along the length of the graphene sheet for pristine, edge functionalized and basal plane functionalized graphene. We have divided the graphene (100 Å long) into 50 bins. For the basal plane case, a sudden temperature drop at the functionalized sites is observed, indicating a reduction in thermal conductivity due to phonon scattering at functionalization sites. However, temperature distribution within the edge functionalized graphene remains linear due to the minimal distortion and shows no such sudden temperature drops. These results clearly demonstrate that edge functionalized graphene has superior heat conduction ability compared to basal plane functionalized graphene.

While above results demonstrate advantage of edge bonding in heat conduction in individual graphene sheets, we next demonstrate that interface thermal conductance is also higher at an individual junction (between polymer chain and graphene) located on edge of graphene as opposed to on the basal plane. This implies that heat conducts into graphene more efficiently through a junction located on edge compared to basal plane. Molecular dynamics simulations were used to compare interface thermal conductance for junction at an edge relative to on the basal plane. Atomic structures for the two cases are shown in Fig. 7a which shows polymer chains bonded on the edge as well as on the basal plane of graphene. Temperature difference of 20 K was applied across the location of bonding (junction) and resulting heat flux was computed. MD computations reveal almost 40% higher heat flux for edge case relative to basal plane case (Fig. 7b). To understand the cause of this higher interface thermal conductance, we performed first-principles Green's function calculations, described next.

## 2.2 Atomistic Green's Function Computations

First-principles atomistic Green's function (AGF)[42,43] method offers several unique advantages for computation of interface conductance such as allowing use of accurate interatomic force interactions derived from density-functional theory as well as use of exact interfacial atomic arrangement. The method further enables detailed microscopic understanding of interfacial thermal transport in terms of transmission of individual phonon modes across the junction. For AGF calculations, system is divided into left contact lead, central interface region and right contact lead. In this work, left lead was comprised of polyethylene chain, right lead was graphene nanoribbon and central region was comprised of the junction between polymer and graphene. Phonon interface thermal conductance is computed using Landauer formalism[44] from the estimate of phonon transmission rates. For this work, second-order interatomic force constants needed for the computation of phonon transmission were derived *ab initio* using open source DFT package QUANTUM-ESPRESSO[35]. Fig. 7c shows that AGF method also predicts almost 60% higher interface thermal conductance for edge case relative to basal plane bonding at 300 K. This agrees qualitatively with MD simulation results.

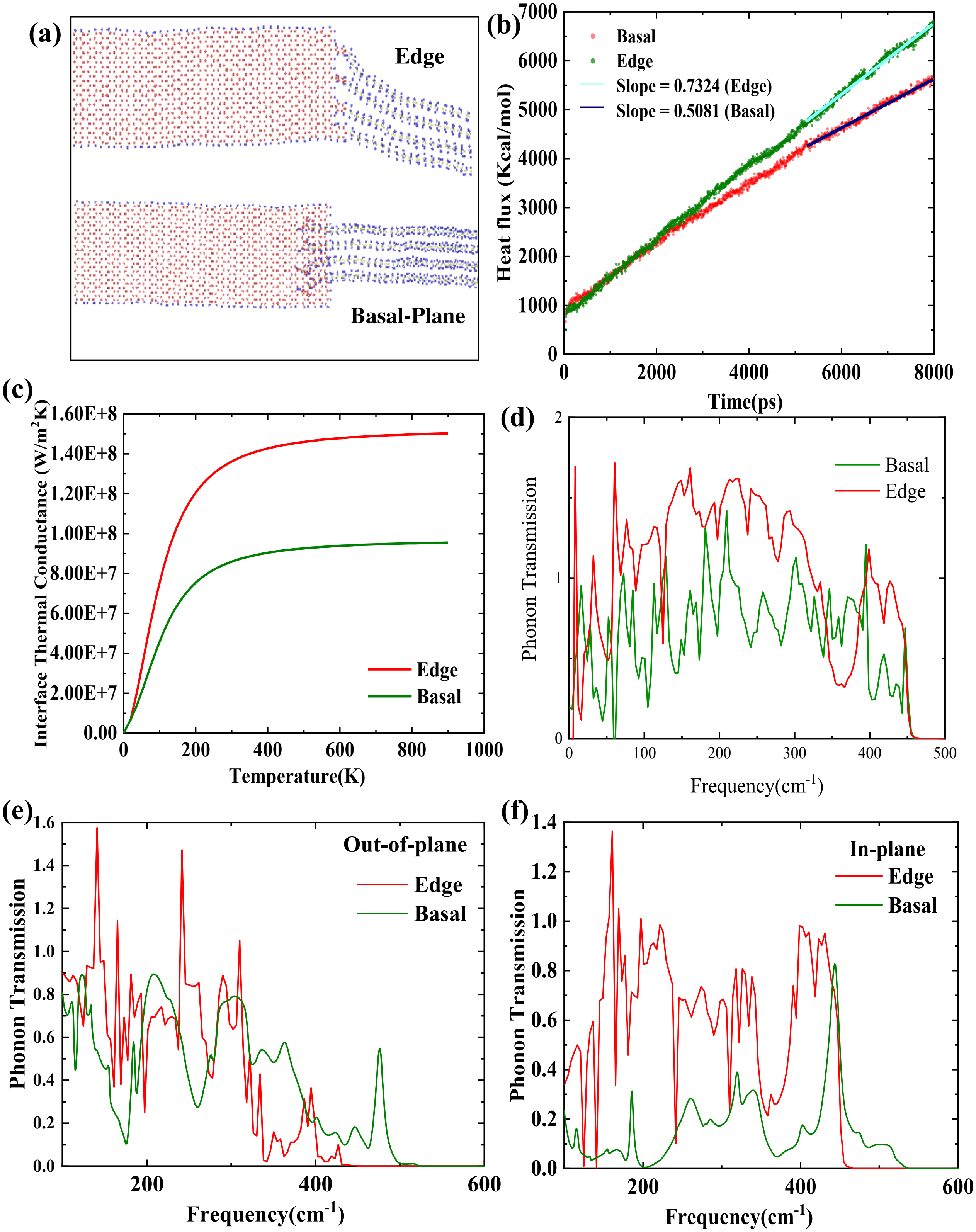

**Figure 7.** a) Relaxed structures of single sheet edge and basal plane functionalization, (b) Heat flux, c) interface thermal conductance, d) overall phonon transmission, e) in-plane phonon transmission, and f) out of-plane phonon transmission of edge and basal plane functionalization.

Fig. 6d shows that this higher thermal conductance is due to the higher phonon transmission for a junction on edge of graphene relative to basal plane. To understand this higher phonon transmission for edge case, we split the total phonon transmission from polymer chain into in-plane and out-of-plane vibration modes of graphene using polarization dependent atomistic Green's function method (details provided in Methods). Results show that transmission from polymer to in-plane vibration modes of graphene is much larger for edge case relative to basal plane bonding. The contribution of out-of-plane vibration modes to total transmission is found to be comparable for edge and basal plane bonding cases. This higher transmission to in-plane phonons of graphene for edge case leads to the higher overall phonon transmission for edge relative to basal plane cases, resulting in higher interfacial thermal conductance for edge case.

Above computations provide a comprehensive understanding of the higher polymer-graphene nanocomposite thermal conductivity through edge relative to basal plane functionalization.

## 3. CONCLUSION

In summary, this work provides a detailed comparison of edge and basal plane functionalization of graphene nanoplatelets in enhancing thermal conductivity of polymer-graphene nanocomposite, through both molecular dynamics study and Green's function calculations. Edge bonding is shown to be capable of leading to superior enhancement in composite thermal conductivity. Molecular dynamics simulations reveal this advantage of edge bonding to be due to all sheets of graphene nanoplatelets interacting efficiently with surrounding polymer through edge bonding. Coupling of the high in-plane thermal conductivity of graphene (~2000 W/mK) with the surrounding polymer through edge bonding leads to an efficient thermal conduction pathway through the composite. For basal plane bonding, only the outermost layers (surface layers) of the nanoplatelet have significant thermal interaction with the polymer. Poor out-of-plane thermal conductivity of graphene (~ 10 W/mK) causes inefficient heat conduction from outer to inner layers, resulting in the overall nanoplatelet being less effective in heat conduction for basal plane bonding compared to edge bonding. Overall, simulations predict 48% higher $k$ through edge relative to basal plane bonding, at 35 weight% composition and for 10 layer thick nanoplatelets. Simulations revealed other advantages of edge bonding, such as lower damping of vibrations in all layers of the nanoplatelet through edge bonding, high thermal conductivity of individual graphene layers through minimal distortion of graphene structure induced by edge functionalization, and higher thermal conductance of individual junction at edge relative to basal plane, mediated by higher phonon transmission to in-plane phonons of graphene. Through detailed study of efficient functionalization schemes, this work opens up new avenues to achieve higher thermal conductivity of polymer-graphene nanocomposites, with important applications in a wide range of thermal management technologies.

**Conflicts of Interest**

There are no conflicts of interest to declare.

**Acknowledgements**

RM, FT, SD and JG acknowledge support from National Science Foundation CAREER award under Award No. #1847129. We also acknowledge the University of Oklahoma Supercomputing Center for Education and Research (OSCER) for providing computing resources for this work.